# A New Type of Compositionally Complex $M_5Si_3$ Silicides: Cation Ordering and Unexpected Phase Stability

Sashank Shivakumar [#], Mingde Qin [#], Dawei Zhang, Chongze Hu, Qizhang Yan, Jian Luo [*]

Department of NanoEngineering; Program of Materials Science and Engineering, University of California San Diego, La Jolla, CA, 92093, USA

## Abstract

A new type of compositionally complex (medium- or high-entropy) $M_5Si_3$ silicides is synthesized. Both $(V_{1/5}Cr_{1/5}Nb_{1/5}Ta_{1/5}W_{1/5})_5Si_3$ and $(Ti_{1/5}Zr_{1/5}Nb_{1/5}Mo_{1/5}Hf_{1/5})_5Si_3$ form single-phase homogenous solid solutions. Notably, $(V_{1/5}Cr_{1/5}Nb_{1/5}Ta_{1/5}W_{1/5})_5Si_3$ forms the hexagonal $\gamma$ ($D8_8$) phase, while all its five constituent binary silicides, $V_5Si_3$, $Cr_5Si_3$, $Nb_5Si_3$, $Ta_5Si_3$, and $W_5Si_3$, are stable in the tetragonal $\alpha$ ($D8_l$) or $\beta$ ($D8_m$) phases. Annealing at 1600°C demonstrates that this hexagonal $\gamma$ phase is stable. Comparison of the experimental and calculated X-ray diffraction patterns, Rietveld refinements, and analysis of aberration-corrected scanning transmission electron microscopy high-angle annular dark-field images suggest cation ordering, which reduces the configurational entropy. This work expands the field of high-entropy and compositional complex ceramics by not only discovering a new compositional complex silicide phase but also demonstrating the cation ordering and unusual phase stability. These compositionally complex silicides can be combined with refractory high-entropy alloys to make the high-entropy counterparts to the Nb-silicide and Mo-Si-B composites.

# Graphical Abstract

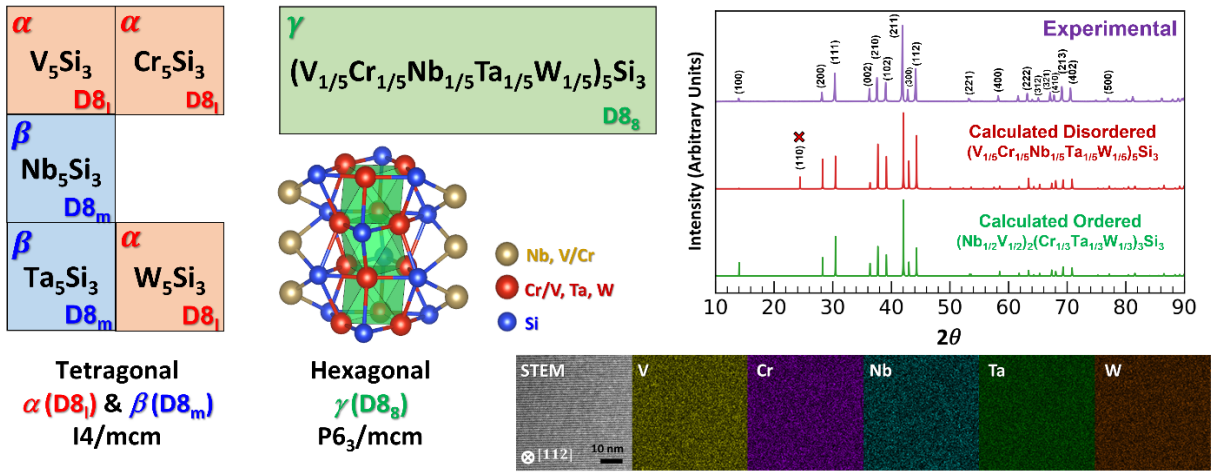

## Highlights:

- A new type of compositionally complex $M_5Si_3$ silicides is synthesized.

- $V_5Si_3$, $Cr_5Si_3$, $Nb_5Si_3$, $Ta_5Si_3$, and $W_5Si_3$ are all stable in tetragonal structures.

- However, $(V_{1/5}Cr_{1/5}Nb_{1/5}Ta_{1/5}W_{1/5})_5Si_3$ forms a hexagonal structure.

- Cation ordering is found, suggesting $[Nb_{1/2}(Cr/V)_{1/2}]_2[(V/Cr)_{1/3}Ta_{1/3}W_{1/3}]_3Si_3$.

- Hexagonal $(Ti_{1/5}Zr_{1/5}Nb_{1/5}Mo_{1/5}Hf_{1/5})_5Si_3$ is also synthesized.

High-entropy alloys (HEAs), which may be more broadly defined as complex concentrated (or compositionally complex) alloys (CCAs), is currently a topic of intense research [1-4]. In comparison with metallic HEAs, the recently developed high-entropy ceramics (HECs) can possess a broad variety of stoichiometries, crystal structures, and bonding characters [5]. For example, various high-entropy oxides (including MO rock-salt [6] and $MO_{2-\delta}$ fluorite [7] oxides with one cation sublattice, as well as $ABO_3$ pervoskite [8] and $A_2B_2O_7$ pyrochlore [9-11] oxides with two cation sublattices), borides (including $MB_2$ [12], MB [13], $M_3B_4$ [14], $MB_4$ [15], and $MB_6$ [16, 17] stoichiometries), rock-salt carbides [18-20] and nitrides [21-23], silicides ($MSi_2$ disilcides) [24, 25], and fluorides ($MF_2$) [26] have been fabricated. Akin to the generalization from HEAs to CCAs, Luo and co-workers proposed in 2020 to broaden HECs to "compositionally complex ceramics (CCCs)", where medium-entropy and non-equimolar compositions can have better properties [5, 11, 27]. In comparison with their metallic counterparts, aliovalent doping, vacancies, and short- and long-range cation (or anion) ordering can be introduced in CCCs as additional routes to tailor their properties.

In 2019, Gild *et al.* [24] and Qin *et al.* [25] independently reported the first synthesis of $(Mo_{0.2}Nb_{0.2}Ta_{0.2}Ti_{0.2}W_{0.2})Si_2$ and $(Ti_{0.2}Zr_{0.2}Nb_{0.2}Mo_{0.2}W_{0.2})Si_2$, respectively, which represent the only class of high-entropy silicides that has been fabricated and characterized to date. In general, metal silicides ($M_xSi_y$) are known for their attractive electrical, thermoelectric, and mechanical properties [28-33]. Specifically, $M_5Si_3$ silicides, such as $Nb_5Si_3$, $Mo_5Si_3$, and $Ti_5Si_3$, are promising high-temperature structural materials, owing to their low densities, high melting points, high-temperature strength, and creep resistance [31-33]. $Ta_5Si_3$ also possesses excellent wear and corrosion resistance [33]. Crucially, $Nb_5Si_3$ and $Mo_5Si_3$ serve as secondary phases to strengthen and enhance the oxidation resistance of Nb-silicide and Mo-Si-B composites [32, 34-36] for potential applications at higher temperatures beyond the capabilities of the Ni-based superalloys. In a study of $HfMo_{0.5}NbTiV_{0.5}Si_x$, Liu *et al* noted the formation of a secondary $(Hf, Nb, Ti)_5Si_3$ phase with the addition of Si to $HfMo_{0.5}NbTiV_{0.5}$ HEAs, which resulted in increased strength and hardness [37]. However, the exact composition and structure of this secondary $M_5Si_3$ phase was not characterized. Here, we report the first synthesis of $(V_{1/5}Cr_{1/5}Nb_{1/5}Ta_{1/5}W_{1/5})_5Si_3$ and $(Ti_{1/5}Zr_{1/5}Nb_{1/5}Mo_{1/5}Hf_{1/5})_5Si_3$ to expand the families of HECs/CCCs, particularly silicide CCCs beyond the high-entropy $MSi_2$ disilicides. Interestingly, we further discovered cation ordering and unexpected phase stability, representing new phenomena in HECs/CCCs that have never been reported previously.

In this study, $(Mo_{0.2}Nb_{0.2}Ta_{0.2}Ti_{0.2}W_{0.2})Si_2$ and $(Ti_{0.2}Zr_{0.2}Nb_{0.2}Mo_{0.2}W_{0.2})Si_2$ were synthesized by mixing elemental powders of Si, Ti, V, Cr, Nb, Mo, Zr, Hf, Ta, and W (>99.5% purity, 325 mesh, Alfa Aesar, MA, USA). For each composition, appropriate amounts of powders were weighed out in 6 g batches. Powders were first vortex mixed for 5 min, and subsequently high energy ball milled (HEBM) in a SPEX 800D mill (SpexCertPrep, NJ, USA) for 50 min. Tungsten carbide lined stainless steel jars and 11.2 mm diameter tungsten carbide milling media were used with a ball-to-powder ratio ~4:1. Milling was carried out under an Argon atmosphere (<15 ppm $O_2$) to reduce oxidation, and 1 wt% (0.06 g) of stearic acid was used as a lubricant. The powders were then loaded into 10 mm graphite dies (Cal Nano, CA, USA) lined with graphite foil in batches of 3 g and subsequently consolidated into dense pellets via spark plasma sintering (SPS) in vacuum ($10^{-3} – 10^{-2}$ torr) using a Thermal Technologies 3000 series SPS machine. The sintering temperature was set to be 1900 °C for $(V_{1/5}Cr_{1/5}Nb_{1/5}Ta_{1/5}W_{1/5})_5Si_3$ and 1800 °C for $(Ti_{1/5}Zr_{1/5}Nb_{1/5}Mo_{1/5}Hf_{1/5})_5Si_3$, respectively. A ramp rate of 100 °C/min was utilized, and a hold time of 20 min was employed under 50 MPa of pressure at isothermal sintering temperatures. Samples were then furnace cooled under vacuum. After sintering, specimens were ground to remove the carbon-contaminated surface layers caused by the graphite tooling, and further ground and polished for characterization. X-ray diffraction (XRD) was performed on a Rigaku Miniflex diffractometer (Cu Kα radiation, 30 kV and 15 mA). Densities were measured by the Archimedes method. The theoretical densities were calculated from ideal stoichiometry and the lattice parameters measured by XRD. Scanning electron microscopy (SEM), energy dispersive X-ray spectroscopy (EDS), and electron backscatter diffraction (EBSD) were conducted using a Thermo-Fisher Apreo SEM equipped with an Oxford N-Max$^N$ EDX detector and an Oxford Symmetry EBSD detector. Annealing was performed at 1600 °C in a tube furnace under flowing Argon. Aberration-corrected scanning transmission electron microscopy (AC STEM) high angular annular dark field (HAADF) imaging and nanoscale elemental mapping were obtained using a JEOL JEM-ARM300F TEM at 300 keV. The TEM lamella was prepared using a Thermo-Fisher Scios DualBeam focused ion beam (FIB). The experimental conditions, nominal *vs.* EDS measured compositions, XRD measured lattice parameters, theoretical, measured, and relative densities, and volume fractions of oxide impurities are summarized in Supplementary Table S1. Specimens of both $(Mo_{0.2}Nb_{0.2}Ta_{0.2}Ti_{0.2}W_{0.2})Si_2$ and $(Ti_{0.2}Zr_{0.2}Nb_{0.2}Mo_{0.2}W_{0.2})Si_2$ achieved >~97.5% relative densities and >~96% phase purities (with 2-4 vol.% oxide impurities).

The density functional theory (DFT) calculations were performed using Vienna Ab initio Simulation Package (VASP) code [34,35]. Perdew-Burke-Ernzerhof (PBE) functional [36] was used to optimize all binary silicide structures. The convergence criteria for all DFT calculations were adopted to be the same as those used in the Materials Project [38].

XRD patterns of the as-synthesized $(V_{1/5}Cr_{1/5}Nb_{1/5}Ta_{1/5}W_{1/5})_5Si_3$ and $(Ti_{1/5}Zr_{1/5}Nb_{1/5}Mo_{1/5}Hf_{1/5})_5Si_3$ are shown in Fig. 1, where the calculated XRD patterns assuming random cation distributions are also shown for comparison. For $(V_{1/5}Cr_{1/5}Nb_{1/5}Ta_{1/5}W_{1/5})_5Si_3$, all the peaks in the experimental XRD patterns in Fig. 1(a) can be indexed to the $Mn_5Si_3$ prototyped hexagonal γ phase (albeit a few missing peaks, which can be explained from cation ordering in the discussion later). A minor amount of $(Zr, Hf)O_2$ impurity was found in $(Ti_{1/5}Zr_{1/5}Nb_{1/5}Mo_{1/5}Hf_{1/5})_5Si_3$, while all other major peaks in Fig. 1(b) for can be indexed to the hexagonal γ phase. Fig. 2(a) is the STEM HAADF image taken along the [112] zone axis, which elucidates the hexagonal structure. The homogeneous distribution of elements was verified for $(V_{1/5}Cr_{1/5}Nb_{1/5}Ta_{1/5}W_{1/5})_5Si_3$ by both microscale SEM-EDS and nanoscale STEM-EDS elemental mapping as shown in Fig. 2(b) and 2(c). Additional SEM micrographs and EDS elemental maps of both compositions are shown in Supplementary Fig. S3, which further confirmed the compositional homogeneity. Minor dark spots in Fig. 2(c) correspond to ~3-4 vol.% secondary amorphous $SiO_2$ phase, which was verified in EDS Si and O maps (Supplementary Fig. S4). The measured grain sizes were 21.6 ± 11.7 μm for $(V_{1/5}Cr_{1/5}Nb_{1/5}Ta_{1/5}W_{1/5})_5Si_3$ and 3.20 ± 1.70 μm for $(Ti_{1/5}Zr_{1/5}Nb_{1/5}Mo_{1/5}Hf_{1/5})_5Si_3$, respectively (based on EBSD maps shown in Supplementary Fig. S5).

Notably, $M_5Si_3$ can form three different crystal structures: the tetragonal α phase ($Cr_5Si_3$-prototyped $D8_l$ structure) and β phase ($W_5Si_3$-prototyped $D8_m$ structure), and the hexagonal γ phase ($Mn_5Si_3$-prototyped $D8_8$ structure); the detailed crystal structures are shown in Supplementary Fig. S8. To examine the phase formation and stability, we conducted DFT calculations of the formation energies for α, β, and γ phases for all the relevant binary $M_5Si_3$ silicides at 0 K and the DFT results are shown in Supplementary Table S2), and we also obtained the equilibrium phase formation from the NIST phase diagrams. Fig. 3 summarizes the phase stability of $(V_{1/5}Cr_{1/5}Nb_{1/5}Ta_{1/5}W_{1/5})_5Si_3$ and $(Ti_{1/5}Zr_{1/5}Nb_{1/5}Mo_{1/5}Hf_{1/5})_5Si_3$, along with their constituent binary silicides, at different temperatures.

Surprisingly, $(V_{1/5}Cr_{1/5}Nb_{1/5}Ta_{1/5}W_{1/5})_5Si_3$ (sintered at 1900°C) formed the hexagonal $\gamma$ phase, while all its five constituent binary silicides ($V_5Si_3$, $Cr_5Si_3$, $Nb_5Si_3$, $Ta_5Si_3$, and $W_5Si_3$) are stable in tetragonal $\alpha$ or $\beta$ phases at both 0 K and high temperatures (except for the decomposition of $Cr_5Si_3$ at 1900°C), as shown in Fig. 3(a). To further test the stability of the $\gamma$ phase, a $(V_{1/5}Cr_{1/5}Nb_{1/5}Ta_{1/5}W_{1/5})_5Si_3$ sample was annealed at 1600°C for 5 h under flowing argon. XRD showed that this unexpected hexagonal $\gamma$ phase was stable during the annealing (other than the formation of a minor W-rich secondary BCC metal phase in the reducing environment; Supplementary Fig. S7). $(Ti_{1/5}Zr_{1/5}Nb_{1/5}Mo_{1/5}Hf_{1/5})_5Si_3$ also formed the hexagonal $\gamma$ phase, which is more expected since three out of its five constituent binary silicides are also stable in the $\gamma$ phase (Fig. 3(b)).

The discrepancies between the experimental XRD patterns and calculated patterns assuming random cation distributions (missing peaks) can be attributed to cation ordering, which is further supported by Rietveld refinements and analysis of the STEM HADDF image. In Fig. 1(a), the (110) peak at ~24º is present in the predicted XRD pattern for disordered $(V_{1/5}Cr_{1/5}Nb_{1/5}Ta_{1/5}W_{1/5})_5Si_3$, but it is absent in the experimentally spectrum. As shown in Supplementary Fig. S8 and Fig. S9, there are two different metal sites with a ratio of 2 : 3 in the hexagonal $\gamma$-$M_5Si_3$, which are denoted as A and B sites; consequently, the chemical formula of $\gamma$-$M_5Si_3$ can be rewritten as $A_2B_3Si_3$. This allows cation ordering, or preferential occupancies of the five metal atoms in A *vs.* B sites.

To further investigate the cation ordering in $(V_{1/5}Cr_{1/5}Nb_{1/5}Ta_{1/5}W_{1/5})_5Si_3$, we first calculated theoretical XRD spectra using VESTA for 10 ordered $A_2B_3Si_5$ structures, where we selected all 10 different combinations of two metals from Nb, Cr, V, Ta, and W to occupy the A sites. As shown in Supplementary Fig. S10, $(Nb_{1/2}V_{1/2})_2(Cr_{1/3}Ta_{1/3}W_{1/3})_3Si_3$ and $(Nb_{1/2}Cr_{1/2})_2(V_{1/3}Ta_{1/3}W_{1/3})_3Si_3$ produced the calculated XRD patterns with all peaks match the experiment perfectly. We further calculated an XRD pattern for $(Nb_{1/2}V_{1/4}Cr_{1/4})_2(V_{1/6}Cr_{1/6}Ta_{1/3}W_{1/3})_3 Si_3$, which can also match all experimental peaks. However, disordered $(V_{1/5}Cr_{1/5}Nb_{1/5}Ta_{1/5}W_{1/5})_5Si_3$ and ordered $(Nb_{1/3}V_{1/3}Cr_{1/3})_2(Nb_{1/3}V_{1/9}Cr_{1/9}Ta_{1/3}W_{1/3})_3Si_3$ produced calculated XRD patterns with extra peaks not found in the experiment.

Furthermore, we conducted Rietveld refinements for the three best fitted ordered structures, along with the disordered structure, and compare them with the measured XRD pattern in Fig. 4(a) (with Rietveld fittings shown in Supplementary Fig. S11). The weighted profile residual ($R_{wp}$) values were 6.62% for $(Nb_{1/2}V_{1/2})_2(Cr_{1/3}Ta_{1/3}W_{1/3})_3Si_3$, 6.75% for $(Nb_{1/2}Cr_{1/2})_2(V_{1/3}Ta_{1/3}W_{1/3})_3Si_3$, and 9.35% for $(Nb_{1/2}V_{1/4}Cr_{1/4})_2(V_{1/6}Cr_{1/6}Ta_{1/3}W_{1/3})_3 Si_3$, respectively, for three ordered structures. In comparison, a larger $R_{wp}$ value of 14.9% was obtained for disordered $(V_{1/5}Cr_{1/5}Nb_{1/5}Ta_{1/5}W_{1/5})_5Si_3$. In addition, clear discrepancies of the (100) and (110) peaks were observed for the disordered case, but not for the three ordered cases. These results suggest cation ordering, with either $(Nb_{1/2}V_{1/2})_2(Cr_{1/3}Ta_{1/3}W_{1/3})_3Si_3$ or $(Nb_{1/2}Cr_{1/2})_2(V_{1/3}Ta_{1/3}W_{1/3})_3Si_3$ being the like structure (with the lowest $R_{wp}$); fully mixing of V and Cr is less likely (because of a high $R_{wp}$ of 9.35%). However, some levels of anti-site cation mixing for V and Cr, as well as other atoms (presumably to less extents), should be inevitable at finite temperatures.

To further verify this proposed cation ordering from fitting XRD patterns, we perform an intensity ratio analysis of the STEM HAADF image shown in Fig. 4(b). Viewed along the [112] zone axis, the $(1\bar{1}0)$ and $(4\bar{4}0)$ planes are composed of B-site atoms, whereas the $(3\bar{3}0)$ plane is composed of the A-site atoms. The ratio of the HAADF intensity of the $(1\bar{1}0)$ *vs.* $(4\bar{4}0) + (3\bar{3}0)$ atomic planes was measured to be ~1.3 from the experimental HAADF image. By assuming the HAADF contrast depends on the square of atomic mass ($Z^2$), this intensity ratio should be 0.667, if metals randomly occupy A and B sites in the disordered $(V_{1/5}Cr_{1/5}Nb_{1/5}Ta_{1/5}W_{1/5})_5Si_3$, which differs substantially from the experimental value of ~1.3. In contrast, this ratio was calculated to be 1.252 for ordered $(Nb_{1/2}V_{1/2})_2(Cr_{1/3}Ta_{1/3}W_{1/3})_3Si_3$ and 1.263 for ordered $(Nb_{1/2}Cr_{1/2})_2(V_{1/3}Ta_{1/3}W_{1/3})_3Si_3$, which match the experimental value of ~1.3 well. In fact, Supplementary Table S5 showed calculated intensity ratios for 13 different combinations, and the best matches are to place Nb and V/Cr on A sites. Noting that V and Cr have similar masses so that we cannot differentiate them in HADDF imaging; they also have similar X-ray scattering capabilities, so that we could not differentiate them via fitting XRD patterns too.

Hence, analyses of XRD patterns and STEM HAADF image intensity both suggest ordered $[Nb(V/Cr)]_2[(Cr/V)TaW]_3Si_3$ structure (albeit some inevitable anti-site disorder due to an entropic effect). Interestingly, this cation ordering reduces the configurational entropy from $ln5 \cdot k_B$ or $\sim 1.61 k_B$ per metal cation to $(\frac{2}{5}ln2 + \frac{3}{5}ln3) \cdot k_B$ or $\sim 0.94 k_B$ per metal cation for perfect ordering. Thus, this composition will be barely "medium-entropy" based on the common definition [3, 5] even with some anticipated anti-site disorder that may increase the configurational entropy to $> 1 k_B$ per metal cation.

The observed cation ordering may be explained based in the atomic/cationic radii of the five elements. Supplementary Fig. S9(b) shows that there are two distinct planes in hexagonal $\gamma$-$M_5Si_3$: (1) a loosely packed plane consisting of only A-site metal atoms, and (2) a close packed plane consisting of both B-site metal and silicon atoms. A prior computational study of alloying $\gamma$-$Nb_5Si_3$ suggested that atoms such as V and Cr with smaller radii (than that of Nb) prefer to occupy the loosely packed plane (A sites), while atoms with a radius larger than Nb, such as Hf and Zr, prefer to occupy the closely packed planes (B sites) [39]. In this case, atoms of larger radii (Ta and W) occupy closely packed B sites, and those with smaller radii (V, Cr, and Nb) occupy loosely packed A sites. However, radii are likely not the only factor; otherwise, $(Cr_{1/2}V_{1/2})_2(Nb_{1/3}Ta_{1/3}W_{1/3})_3Si_3$ should be the most favorable configuration, but this configuration produced the XRD pattern and HADDF image that did not fit the experiments.

It is possible that this cation ordering also stabilize the otherwise unstable hexagonal $\gamma$ phase. In fact, the prior DFT study suggested that alloying $\beta$-$Nb_5Si_3$ with V, Cr, Mo, and W can preferentially stabilize the hexagonal $\gamma$ phase [39].

Similar analyses of XRD patterns and Rietveld refinements for $(Ti_{1/5}Zr_{1/5}Nb_{1/5}Mo_{1/5}Hf_{1/5})_5Si_3$ (Supplementary Figs. S14 and S15) also suggested cation ordering. However, we are unable to determine the exact cation configurations. It is possible more disordered, but the fully disordered structure did not fit the experiment. It is less a surprise that $(Ti_{1/5}Zr_{1/5}Nb_{1/5}Mo_{1/5}Hf_{1/5})_5Si_3$ also formed the hexagonal $\gamma$ phase, since three out of the five binary silicides ($Ti_5Si_3$, $Zr_5Si_3$, and $Hf_5Si_3$) are stable in the $\gamma$ phase (Fig. 3(b)).

In summary, we have fabricated a new type of compositionally complex $M_5Si_3$ silicides, including $(V_{1/5}Cr_{1/5}Nb_{1/5}Ta_{1/5}W_{1/5})_5Si_3$ and $(Ti_{1/5}Zr_{1/5}Nb_{1/5}Mo_{1/5}Hf_{1/5})_5Si_3$, for the first time. Notably, $(V_{1/5}Cr_{1/5}Nb_{1/5}Ta_{1/5}W_{1/5})_5Si_3$ is stable in the hexagonal $\gamma$ phase, while all its five constituent binary silicides, $V_5Si_3$, $Cr_5Si_3$, $Nb_5Si_3$, $Ta_5Si_3$, and $W_5Si_3$, are stable in the tetragonal α or β phases. Analyses of XRD patterns (including Rietveld refinements) and the AC STEM HAADF image suggested cation ordering, which reduces the configurational entropy but may stabilize the otherwise unstable $\gamma$ phase. This work expands the field of HECs and CCCs by not only discovering a new compositional complex silicide phase but also demonstrating the cation ordering and unusual phase stability, which represent new phenomena in HECs/CCCs that have not been reported previously.

These new compositionally complex silicides can serve as secondary phases in refractory HEAs to not only strengthen them, but also enhance their oxidation resistance. In other words, we envision that these new compositionally complex silicides can be composited with BCC-based refractory HEAs to make high-entropy counterparts to the Nb-silicide and Mo-Si-B composites [32, 34-36] for high-temperature applications beyond the current superalloys.

**Acknowledgement:** This research was supported by the National Science Foundation (NSF) Materials Research Science and Engineering Center program through the UC Irvine Center for Complex and Active Materials (DMR-2011967). This work utilized the shared facilities at the San Diego Nanotechnology Infrastructure of UCSD, a member of the National Nanotechnology Coordinated Infrastructure (supported by the NSF ECCS-1542148), and the Irvine Materials Research Institute (also supported in part by the NSF DMR-2011967).

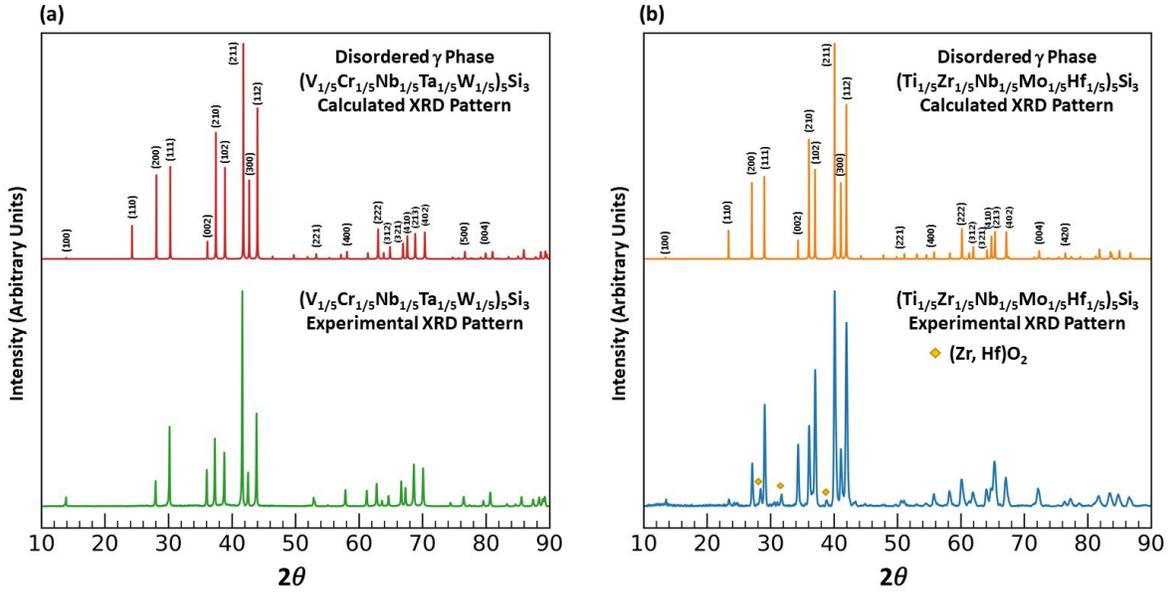

**Figure 1.** Indexed theoretical *vs.* experimental XRD patterns of (a) $(V_{1/5}Cr_{1/5}Nb_{1/5}Ta_{1/5}W_{1/5})_5Si_3$ and (b) $(Ti_{1/5}Zr_{1/5}Nb_{1/5}Mo_{1/5}Hf_{1/5})_5Si_3$. Here, the theoretical XRD patterns were calculated based on the disordered $\gamma$ phase, while assuming random cation distributions on all metal sites. All peaks in the experimental XRD patterns in (a) can be indexed to the hexagonal $\gamma$ phase, although there are some missing peaks in comparison with the calculated XRD pattern of the disordered the $\gamma$ phase. There is a minor amount of impurity in (b) that can be indexed to $(Zr, Hf)O_2$; all other peaks can be indexed to the hexagonal $\gamma$ phase.

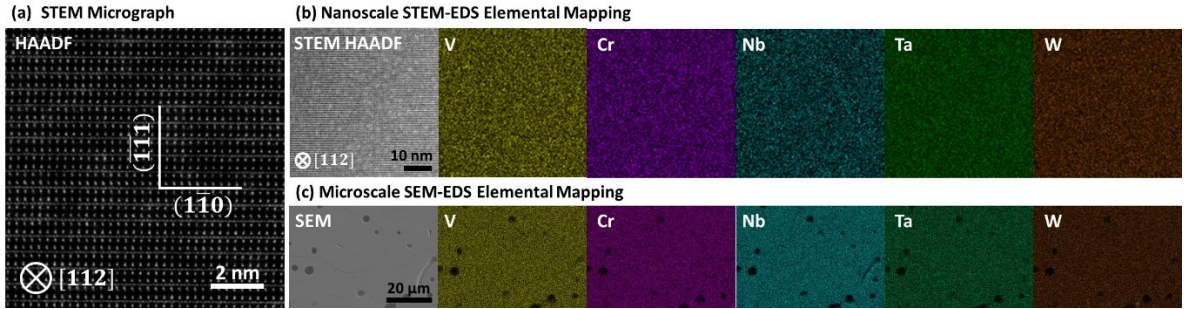

**Figure 2.** (a) STEM HAADF image of single-phase $(V_{1/5}Cr_{1/5}Nb_{1/5}Ta_{1/5}W_{1/5})_5Si_3$ with hexagonal structure (γ phase), viewed along the [112] zone axis. (b) STEM micrograph and corresponding EDS elemental maps at the nanoscale. (c) SEM micrograph and corresponding EDS elemental maps at the microscale. The compositions are homogenous at both nanoscale and microscale. Minor dark spots in (c) corresponds to ~3-4 vol. % secondary amorphous $SiO_2$ phase, which was verified in EDS Si and O maps shown in Supplementary Fig. S4.

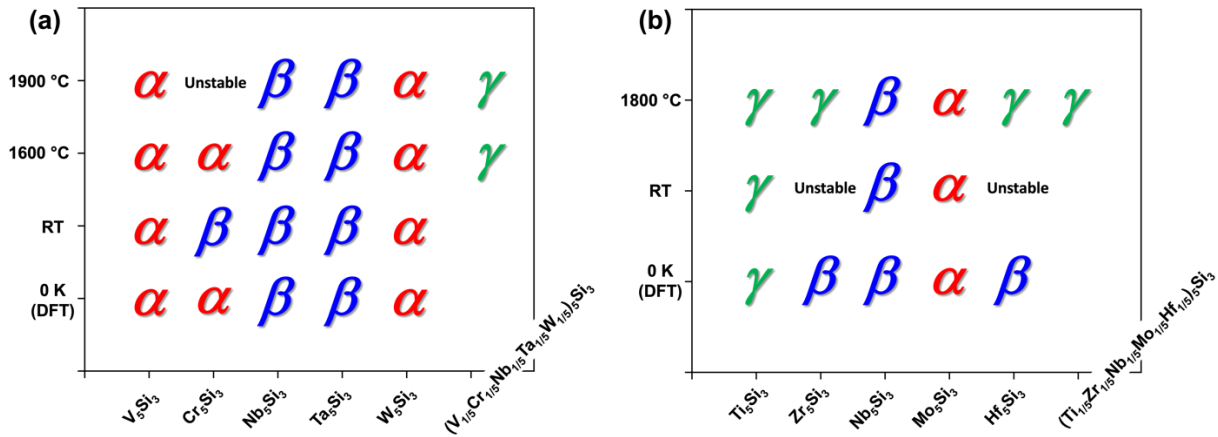

**Figure 3.** Phase stability maps of (a) $(V_{1/5}Cr_{1/5}Nb_{1/5}Ta_{1/5}W_{1/5})_5Si_3$ and (b) $(Ti_{1/5}Zr_{1/5}Nb_{1/5}Mo_{1/5}Hf_{1/5})_5Si_3$ along with their constituent binary silicides, including the 0 K DFT predicted phases (see Supplementary Table 1 for the DFT calculated formation energies) and the stable phases at room temperature, the synthesis temperature of 1900 ºC or 1800 ºC, and the annealing temperature of 1600 ºC. Notably, $(V_{1/5}Cr_{1/5}Nb_{1/5}Ta_{1/5}W_{1/5})_5Si_3$ forms the hexagonal $\gamma$ phase, while all its binary silicides, $V_5Si_3$, $Cr_5Si_3$, $Nb_5Si_3$, $Ta_5Si_3$, and $W_5Si_3$, are stable in tetragonal $\alpha$ or $\beta$ structures (but not in the hexagonal $\gamma$ phase at any of these temperatures). The formation of $(Ti_{1/5}Zr_{1/5}Nb_{1/5}Mo_{1/5}Hf_{1/5})_5Si_3$ in the hexagonal $\gamma$ phase is more expected since three out of five its binary silicides are also stable in the $\gamma$ phase at the synthesis temperature.

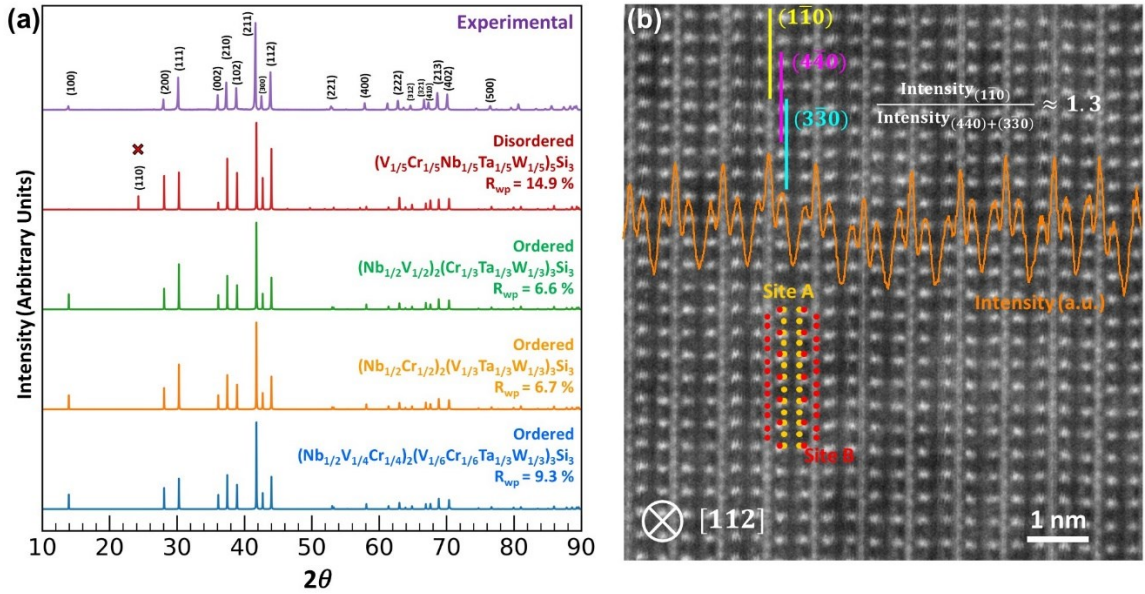

**Figure 4.** (a) Measured XRD pattern for $(V_{1/5}Cr_{1/5}Nb_{1/5}Ta_{1/5}W_{1/5})_5Si_3$ and calculated XRD patterns for several selected cases of cation configurations (for occupying the A *vs.* B sites in $\gamma$-$A_2B_3Si_5$), including disordered and three cases of ordered configurations. The $R_{wp}$ values are the fitting residuals from Rietveld refinements (Supplementary Fig. S7). (b) An intensity ratio analysis performed on a raw STEM HAADF image along the [112] zone axis. The atomic planes $(1\bar{1}0)$ and $(4\bar{4}0)$ are composed of B-site atoms, whereas the atomic plane $(3\bar{3}0)$ is composed of the A-site atoms. The ratio of the HAADF intensity of the $(1\bar{1}0)$ *vs.* $(4\bar{4}0) + (3\bar{3}0)$ atomic planes was determined to be ~1.3 by digital image processing. Supplementary Table S5 tabulates the calculated theoretical intensity ratios based on site A and site B occupancy. This intensity ratio should be ~0.67 for the disordered $(V_{1/5}Cr_{1/5}Nb_{1/5}Ta_{1/5}W_{1/5})_5Si_3$ (where all metals randomly occupy A and B sites), which differs from the experimental value of ~1.3. In contrast, this ratio is ~1.25 for ordered $(Nb_{1/2}V_{1/2})_2(Cr_{1/3}Ta_{1/3}W_{1/3})_3Si_3$ and ~1.26 for ordered $(Nb_{1/2}Cr_{1/2})_2(V_{1/3}Ta_{1/3}W_{1/3})_3Si_3$, which match the experimental value of ~1.3. Thus, analyses of both XRD patterns and the STEM HAADF image intensity suggest ordered $[Nb(V/Cr)]_2[(Cr/V)TaW]_3Si_3$ cation configuration (albeit inevitable anti-site disorder). It is noted that we cannot differentiate V *vs.* Cr because of their similar $Z$ numbers and X-ray scattering factors.